\documentclass[preprint,showpacs,preprintnumbers,amsmath,amssymb,12pt]{revtex4-1}
\usepackage{lipsum}
\usepackage{setspace}
\usepackage{bm}
\usepackage[colorlinks,linkcolor = black,anchorcolor = blue,urlcolor = blue,	citecolor = blue]{hyperref}
\usepackage{float}
\usepackage{graphicx}
\usepackage{makecell}
\usepackage{lineno}
\usepackage{siunitx}
\usepackage{multirow}
\usepackage[utf8]{inputenc}
\DeclareUnicodeCharacter{2009}{\,}   
\DeclareUnicodeCharacter{2212}{-}    

\begin{document}

\title{Doping Evolution of Nodal Electron Dynamics in Trilayer Cuprate Superconductor Bi$_2$Sr$_2$Ca$_2$Cu$_3$O$_{10+\delta}$ Revealed by Laser-Based Angle-Resolved Photoemission Spectroscopy}

\author{Hao Chen$^{1,2,\dag}$, Jumin Shi$^{1,2,\dag}$, Xiangyu Luo$^{1}$, Yinghao Li$^{1,2}$, Yiwen Chen$^{1,2}$, Chaohui Yin$^{1}$, Yingjie Shu$^{1,2}$, Jiuxiang Zhang$^{1,2}$, Taimin Miao$^{1,2}$, Bo Liang$^{1,2}$, Wenpei Zhu$^{1,2}$, Neng Cai$^{1,2}$, Xiaolin Ren$^{1,2}$, Chengtian Lin$^{3}$, Shenjin Zhang$^{4}$, Zhimin Wang$^{4}$, Fengfeng Zhang$^{4}$, Feng Yang$^{4}$, Qinjun Peng$^{4}$, Zuyan Xu$^{4}$, Guodong Liu$^{1,2,5}$, Hanqing Mao$^{1,2,5}$, Xintong Li$^{1,2,5}$, Lin Zhao$^{1,2,5,*}$ and X. J. Zhou$^{1,2,5,*}$}

\affiliation{
   \\$^{1}$Beijing National Laboratory for Condensed Matter Physics, Institute of Physics, Chinese Academy of Sciences, Beijing 100190, China.
   \\$^{2}$University of Chinese Academy of Sciences, Beijing 100049, China.
   \\$^{3}$Max Planck Institute for Solid State Research, Heisenbergstrasse 1, D-70569 Stuttgart, Germany.
   \\$^{4}$Technical Institute of Physics and Chemistry, Chinese Academy of Sciences, Beijing 100190, China.
   \\$^{5}$Songshan Lake Materials Laboratory, Dongguan, Guangdong 523808, China.
   \\$^{\dag}$These authors contributed equally to this work.
   \\$^{*}$Corresponding author: LZhao@iphy.ac.cn, XJZhou@iphy.ac.cn
}

\date{\today}
\maketitle

\begin{center}
  \textbf{ABSTRACT}
\end{center}
\vspace{1em}

The doping evolution of the nodal electron dynamics in the trilayer cuprate superconductor $\mathrm{Bi_{2}Sr_{2}Ca_{2}Cu_{3}O_{10+\delta}}$ (Bi2223) is investigated using high-resolution laser-based angle-resolved photoemission spectroscopy (ARPES). Bi2223 single crystals with different doping levels are prepared by controlled annealing which cover the underdoped, optimally-doped and overdoped regions. The electronic phase diagram of Bi2223 is established which describes the T$_c$ dependence on the sample doping level. The doping dependence of the nodal Fermi momentum for the outer (OP) and inner (IP) CuO$_2$ planes is determined. Charge distribution imbalance between the OP and IP CuO$_2$ planes is quantified, showing enhanced disparity with increasing doping. Nodal band dispersions demonstrate a prominent kink at $\sim$94\,meV in the IP band, attributed to the unique Cu coordination in the IP plane, while a weaker $\sim$60\,meV kink is observed in the OP band. The nodal Fermi velocity of both OP and IP bands is nearly constant at $\sim$1.62\,eV\AA\;independent of doping. These results provide important information to understand the origin of high T$_c$ and superconductivity mechanism in high temperature cuprate superconductors.


\vspace{3mm}

\begin{figure*}[tbp]
\begin{center}
\includegraphics[width=1\textwidth,angle=0]{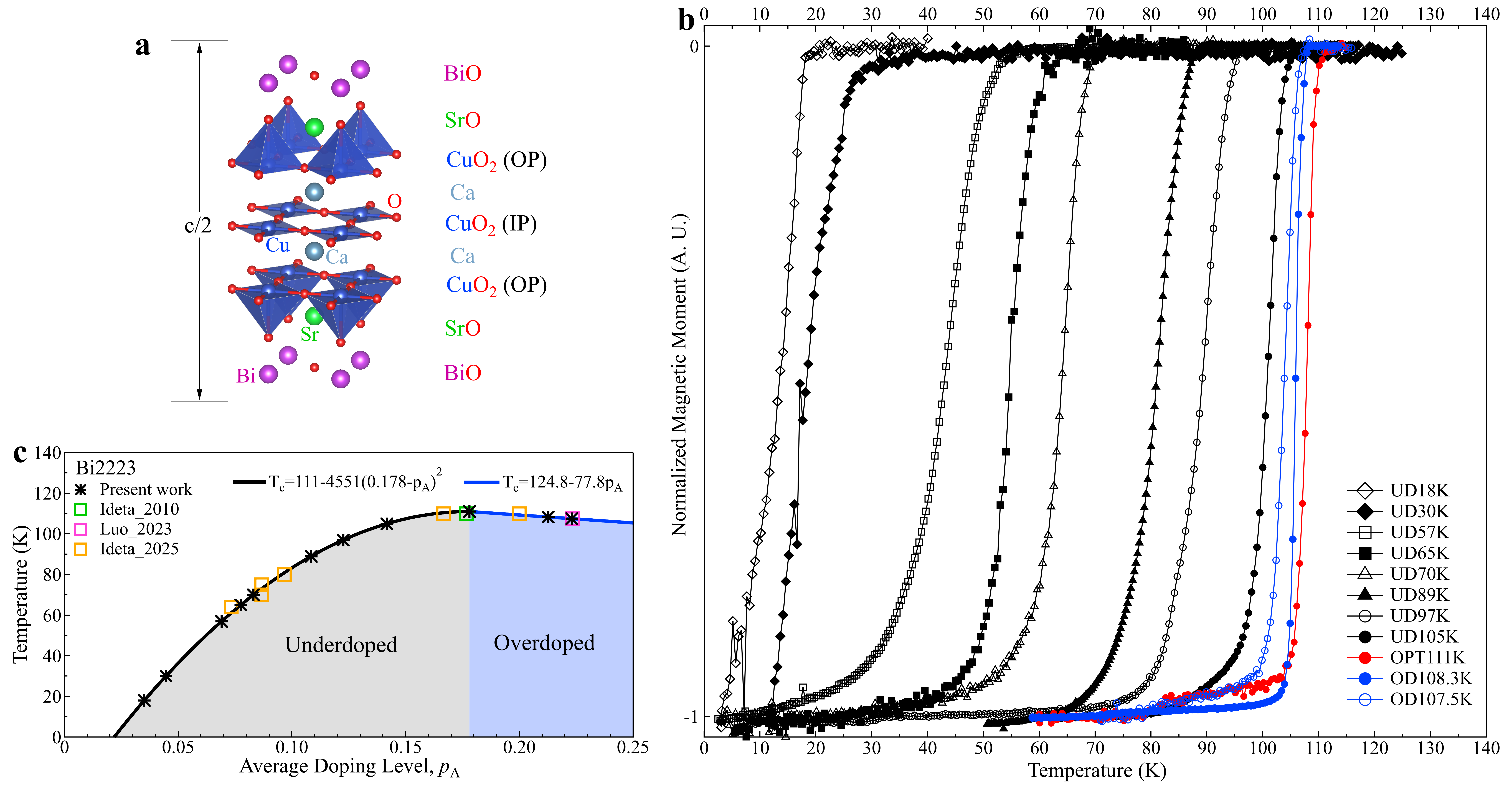}
\end{center}
\caption{{\bf Preparation of Bi2223 single crystals with different doping levels and the related phase diagram.}
\textbf{a}, Schematic crystal structure of Bi2223. Here only half of the unit cell is drawn and the complete unit cell consists of two such structural units displaced with each other by (a/2, b/2). Each structural unit contains three adjacent CuO$_2$ planes with one inner plane (IP) and two outer planes (OPs).
\textbf{b}, AC magnetic measurements of Bi2223 single crystals with different doping levels and different T$_c$s. The optimally-doped Bi2223 single crystals have a $T_c$ at 111\,K. The optimally-doped samples were annealed in vacuum to obtain underdoped samples with $T_c$s at 105\,K, 97\,K, 89\,K, 70\,K, 65\,K, 57\,K, 30\,K and 18\,K. The optimally-doped samples were also annealed under high oxygen pressure of $\sim$170 atmospheres to obtain overdoped samples with $T_c$s at 108.3\,K and 107.5\,K.
\textbf{c}, Phase diagram of Bi2223. The underdoped region is obtained by fitting the data of T$_c$s and doping levels from previous studies with a parabolic form while the overdoped region is obtained by fitting the previous data of T$_c$s and doping levels with a linear form\cite{ideta_enhanced_2010, luo_electronic_2023, ideta_proximity-induced_2025}. Here the doping level $p_A$ refers to the average value of doping levels on all the three CuO$_2$ planes ($p_C$). The doping levels of our samples with different T$_c$s are determined based on this phase diagram.
}
\label{Tc}
\end{figure*}

\clearpage
The trilayer cuprates (Fig. \ref{Tc}a) are unique among all the discovered cuprate superconductors\cite{chu_hole-doped_2015}. First, it has been found that the superconducting transition temperature (T$_c$) depends sensitively on the number of CuO$_2$ planes (n) per structural unit and the trilayer cuprates give the highest T$_c$ among the same family\cite{scott1994layer, chakravarty2004explanation, iyo2007t, eisaki2004effect, chu_hole-doped_2015}. Second, the trilayer cuprates exhibit an unusual electronic phase diagram distinct from that established for the single-layer and bilayer cuprates\cite{keimer_quantum_2015}. In the general phase diagram, T$_c$ goes up with increasing doping in the underdoped region, reaches a maximum at the optimal doping, and then decreases to zero in the overdoped region. But in the trilayer cuprate $\mathrm{Bi_{2}Sr_{2}Ca_{2}Cu_{3}O_{10+\delta}}$ (Bi2223), while T$_c$ increases with increasing doping in the underdoped region, it stays nearly constant in the overdoped region\cite{fujii_doping_2002, piriou2008effect}. The electronic structure studies of the trilayer cuprates are important to understand the origin of the high T$_c$ and the superconductivity mechanism. However, due to the sample limitation, the angle-resolved photoemission (ARPES) measurements on Bi2223 are rather scarce and mostly focused on the optimally-doped samples\cite{feng_electronic_2002, ideta_angle-resolved_2010, ideta_enhanced_2010, ideta_energy_2012, ideta_effect_2013, kunisada_observation_2017, ideta_hybridization_2021, luo_electronic_2023, ideta_proximity-induced_2025}.

In this paper, we prepared a series of Bi2223 single crystal samples with various doping levels by annealing the samples under different conditions. We carried out high-resolution laser-based ARPES measurements on the nodal electronic structures of these samples. The doping dependence of the charge distribution among the three CuO$_2$ planes is determined in Bi2223. The evolution of the nodal electron dynamics with doping is revealed.

High-resolution ARPES measurements were carried out using our lab-based ARPES system equipped with a 6.994\,eV vacuum ultraviolet (VUV) laser and an angle-resolved time-of-flight (ARToF) electron energy analyzer, as well as another laser-based ARPES setup utilizing a 6.994\,eV VUV laser and a DA30L hemispherical electron energy analyzer (Scienta-Omicron)\cite{liu_development_2008, zhou_new_2018}. The ARToF ARPES system enables simultaneous acquisition of photoelectrons across two-dimensional momentum space ($k_x$, $k_y$). For both ARPES systems, the energy resolution was set at $\sim$1\,meV, and the angular resolution was approximately 0.3\,degrees. All Bi2223 samples were cleaved \textit{in situ} at low temperature and measured in ultrahigh vacuum with a base pressure better than $5\times10^{-11}$\,Torr. The Fermi level is referenced by measuring a clean polycrystalline gold which is electrically connected to the sample. 

Bi2223 single crystals were grown by the traveling-solvent floating-zone method\cite{lin_growth_2002, liang_single_2002}. The samples were characterized by AC magnetic measurements (Fig. \ref{Tc}b) and T$_c$ is deﬁned by the onset of the Meissner effect. The optimally-doped (OPT) samples were obtained by annealing the as-grown Bi2223 single crystals in air and they have a T$_c$ at 111\,K. Then the optimally-doped Bi2223 samples were annealed under two different conditions to adjust their doping levels. To get the underdoped (UD) samples, the optimally-doped samples were annealed in vacuum ($1\times10^{-3}$\,mbar) at different temperatures for different durations\cite{lin_growth_2002, liang_single_2002, shimizu_crystal_2002, fujii_doping_2002, piriou2008effect, wei_preparation_2010}. As listed in Table \ref{2223} and shown in Fig. \ref{Tc}b, the underdoped, heavily underdoped and even non-superconducting Bi2223 samples can be obtained. To reach the overdoped (OD) region, the optimally-doped samples were annealed under high oxygen pressure ($\sim$170 atmospheres) at various temperatures for different durations\cite{shimizu_crystal_2002, fujii_doping_2002, piriou2008effect, luo_electronic_2023}, as shown in Fig. \ref{Tc}b. For the overdoped samples, we find that T$_c$ is nearly a constant; it only slightly decreases with the increasing doping. The most overdoped Bi2223 sample has a T$_c$ at 107.5\,K. For convenience, we name our Bi2223 samples by their doping state (underdoped, optimally-doped and overdoped) and T$_c$. For example, UD57K represents the Bi2223 sample which is underdoped and has a T$_c$ at 57\,K.

Although the T$_c$ dependence on the doping level is qualitatively observed in Bi2223\cite{fujii_doping_2002, piriou2008effect}, the phase diagram is still lacking to quantitatively describe the relation between T$_c$ and the doping level. To establish such a phase diagram, Bi2223 samples covering the underdoped, optimally-doped and overdoped regions are required. Their T$_c$s need to be precisely measured and the corresponding doping levels need to be determined. Based on the previous ARPES measurements on Bi2223\cite{ideta_enhanced_2010, luo_electronic_2023, ideta_proximity-induced_2025}, we established the electronic phase diagram of Bi2223 as shown in Fig. \ref{Tc}c. On the underdoped side, we fitted the data from the underdoped samples\cite{ideta_proximity-induced_2025} (orange squares) and the optimally-doped sample\cite{ideta_enhanced_2010} (green squares) by assuming a parabolic form and get the relation $T_c=111-4551(0.178-p_A)^2$, as shown by the black line in Fig. \ref{Tc}c. On the overdoped side, we fitted the data from the optimally-doped sample\cite{ideta_enhanced_2010} (green squares) and the overdoped sample\cite{luo_electronic_2023} (pink squares) by assuming a linear form and get the relation $T_c=124.8-77.8p_A$, as shown by the blue line in Fig. \ref{Tc}c. By combining the underdoped and overdoped regions, we established the quantitative relation between T$_c$ and the doping level in Bi2223. When T$_c$ of the Bi2223 samples is measured (Fig. \ref{Tc}b), their corresponding doping levels can be determined as shown by asterisks in Fig. \ref{Tc}c for our samples.

\begin{figure*}[tbp]
\begin{center}
\includegraphics[width=1\textwidth,angle=0]{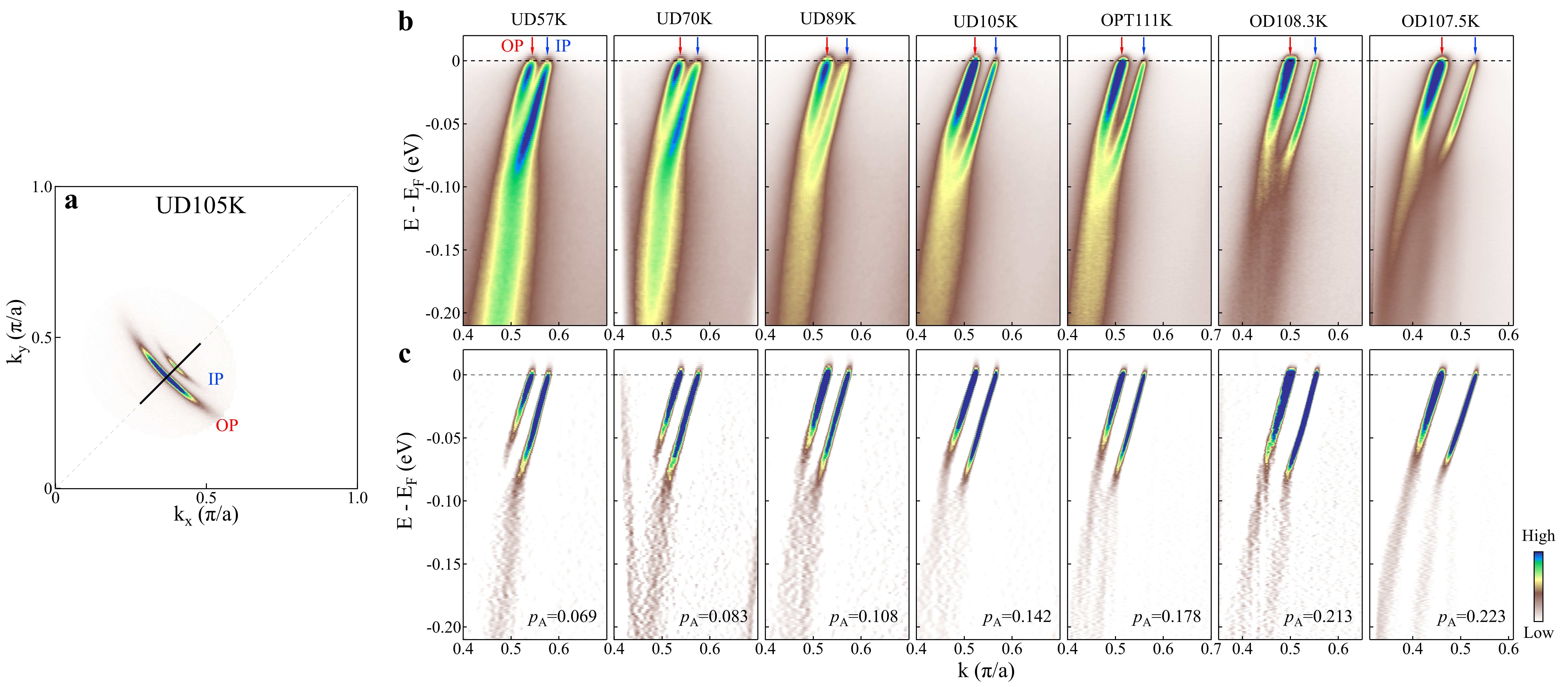}
\end{center}
\caption{{\bf Band structures measured along the nodal direction at 15\,K in Bi2223 samples with different doping levels.}
\textbf{a}. Fermi surface mapping measured in the UD105K Bi2223 sample at 15\,K.
\textbf{b}, Band structures measured along the nodal momentum cut in UD57\,K, UD70\,K, UD89\,K, UD105\,K, OPT111\,K, OD108.3\,K and OD107.5\,K Bi2223 samples. The location of the nodal momentum cut is indicated by the black line in \textbf{a}. The OP and IP bands are marked by red and blue arrows, respectively.
\textbf{c}, Corresponding MDC second-derivative images from \textbf{b}. The doping level (average sample doping $p_A$) of each sample is indicated in each panel.
}
\label{Band}
\end{figure*}

Figure \ref{Band}a shows a typical Fermi surface mapping in the UD105K Bi2223 sample measured at 15\,K by using our ARToF-ARPES system. Since our laser-ARPES measurements can cover only a small portion of the Brillouin zone, the position of the Fermi surface mapping, particularly along the nodal direction, is determined by referring to the previous ARPES measurements on Bi2223 samples\cite{ideta_enhanced_2010, luo_electronic_2023, ideta_proximity-induced_2025}. Two main Fermi surface sheets are clearly observed, labelled as OP and IP in Fig. \ref{Band}a. The OP and IP Fermi surface sheets originate mainly from the outer CuO$_2$ planes and the inner CuO$_2$ plane, respectively. The separation of these two Fermi surface sheets reflects the difference of charge carrier distribution among the OP and IP CuO$_2$ planes. Since the outer CuO$_2$ planes are adjacent to the charge reservoir layers, the doping level of the outer planes is higher than that of the inner plane. This charge distribution imbalance has been observed in the NMR and STEM measurements of the multilayer cuprates\cite{mukuda_high-tc_2012, wang_correlating_2023}. 

Figure \ref{Band}b displays band structures of Bi2223 with different doping levels measured at 15\,K along the nodal momentum cut that is indicated as the black line in Fig. \ref{Band}a. Here the UD57K, UD70K, UD89K, UD105K and OD108.3K Bi2223 samples are measured by using our ARToF-ARPES system while the OPT111K and OD107.5K Bi2223 samples are measured by using our Laser-ARPES system. To identify bands more clearly, Fig. \ref{Band}c shows the corresponding second-derivative images with respect to the momentum obtained from Fig. \ref{Band}b. Two main bands are clearly observed in Fig. \ref{Band}b and Fig. \ref{Band}c, which are labelled as the OP band (red arrow) and the IP band (blue arrow), corresponding to the OP and IP Fermi surface sheets in Fig. \ref{Band}a. The two bands exhibit a systematic evolution with the increasing doping level. First, the separation between the two bands increases with the increase of the doping level. Second, the relative band intensity varies sensitively with the increasing doping level; the OP band get stronger while the IP band becomes weaker. Third, with increasing doping, the OP band extends to higher binding energy and becomes fully visible in the heavily overdoped samples. Fourth, the OP bands become broad in the heavily overdoped samples (OD108.3K and OD107.5K). This is attributed to the enhanced splitting of the OP Fermi surface due to the increased interlayer coupling in the heavily overdoped samples\cite{luo_electronic_2023}.

\begin{figure*}[tbp]
\begin{center}
\includegraphics[width=1\textwidth,angle=0]{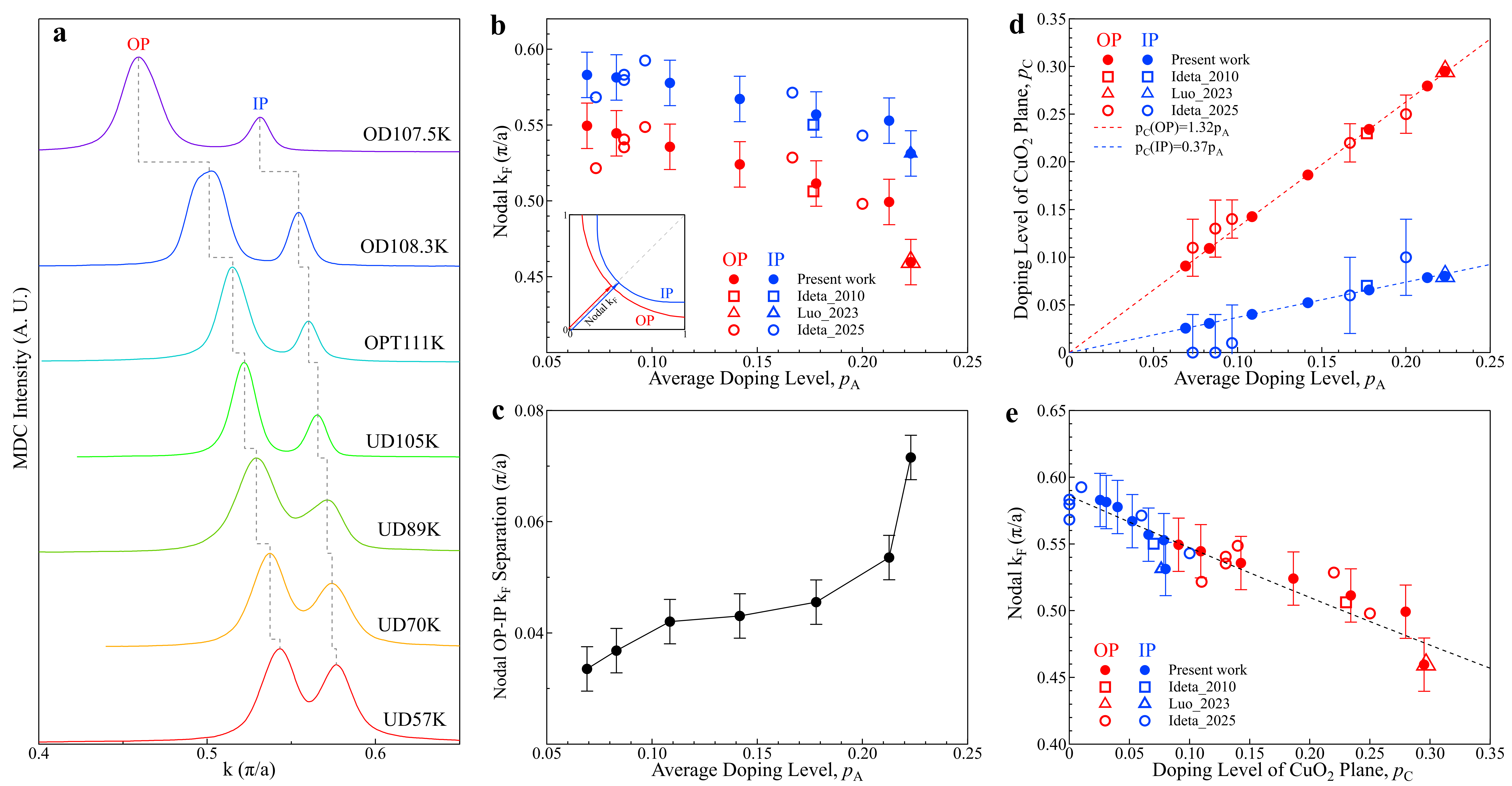}
\end{center}
\caption{{\bf Doping dependence of the nodal Fermi momentum and the charge distribution in OP and IP CuO$_2$ planes in Bi2223.}
\textbf{a}, MDCs at the Fermi level obtained from Fig. \ref{Band}b. 
\textbf{b}, The nodal Fermi momentum $k_F$ of the OP (red symbols) and IP (blue symbols) bands obtained from \textbf{a} as a function of the average doping level ($p_A$). The nodal Fermi momentum $k_F$ is defined in the inset of \textbf{b}.
\textbf{c}, The nodal Fermi momentum difference between the OP and IP bands in our Bi2223 samples.
\textbf{d}, The charge distribution in OP (red symbols) and IP (blue symbols) CuO$_2$ planes in Bi2223 with different doping levels. The doping levels on the OP CuO$_2$ planes from the previous ARPES measurements\cite{ideta_enhanced_2010, luo_electronic_2023, ideta_proximity-induced_2025} are fitted by a linear line (red dashed line). Then the doping levels on the IP CuO$_2$ planes are deduced (blue dashed line). The charge distribution in OP and IP CuO$_2$ planes in our Bi2223 samples are determined based on the two lines.
\textbf{e}, The nodal Fermi momentum $k_F$ of the OP (red symbols) and IP (blue symbols) bands as a function of the doping level of their corresponding CuO$_2$ planes ($p_C$).
}
\label{kf}
\end{figure*}

To quantitatively determine the band position and the OP-IP band separation, Fig. \ref{kf}a presents momentum distribution curves (MDCs) at the Fermi level in Bi2223 samples with different doping levels obtained from Fig. \ref{Band}b. Two main MDC peaks are observed, labelled as OP and IP which correspond to the OP and IP bands in Fig. \ref{Band}b. The positions of these two peaks are presented in Fig. \ref{kf}b, together with the results from the previous ARPES measurements\cite{ideta_enhanced_2010, luo_electronic_2023, ideta_proximity-induced_2025}. The momentum separation between the OP and IP bands is plotted in Fig. \ref{kf}c. The nodal Fermi momentum $k_F$ for both the OP and IP bands decreases monotonically with the increasing average doping level ($p_A$) (Fig. \ref{kf}b). In the meantime, the nodal OP-IP $k_F$ separation increases with the increase of the average doping level ($p_A$) (Fig. \ref{kf}c). With the establishment of these quantitative relations, it is possible to determine the doping level and T$_c$ of the measured area by measuring its nodal band structure in case the Bi2223 sample is inhomogeneous or doped \textit{in situ} by annealing or surface deposition during ARPES measurements.

Since the doped holes are distributed unevenly among three CuO$_2$ planes in Bi2223, it is important to know how the charge carriers are quantitatively allocated among the outer and inner CuO$_2$ planes when the overall doping level changes. Previous ARPES measurements on Bi2223 with different doping levels have provided some information of the charge distribution on the outer and inner CuO$_2$ planes which are plotted in Fig. \ref{kf}d as empty symbols\cite{ideta_enhanced_2010, luo_electronic_2023, ideta_proximity-induced_2025}. The doping level variation of the outer CuO$_2$ plane with the average doping level basically follows a linear line crossing the (0,0) origin which can be described as $p_C(OP)=1.32p_A$ (red dashed line in Fig. \ref{kf}d). Then the doping level variation of the inner CuO$_2$ plane can be derived to be $p_C(IP)=0.37p_A$ (blue dashed line in Fig. \ref{kf}d). With the establishment of these relations, the charge distribution of our measured samples can be determined as shown by solid circles in Fig. \ref{kf}d. 

After the charge distribution on the OP and IP CuO$_2$ planes is known from Fig. \ref{kf}d, we can replot Fig. \ref{kf}b to show the dependence of the nodal Fermi momentum on the doping level of the individual CuO$_2$ plane, as shown in Fig. \ref{kf}e. The nodal Fermi momentum then follows a similar trend for both the OP and IP CuO$_2$ planes. It decreases monotonically with the increasing doping level of the CuO$_2$ plane ($p_C$). As seen in Fig. \ref{kf}d, with increasing average doping level, the doping level difference between the OP and IP CuO$_2$ planes also increases. According to Fig. \ref{kf}e, this will lead to the increase of the nodal OP-IP $k_F$ separation (Fig. \ref{kf}c). Therefore, the increase of the nodal OP-IP $k_F$ separation (Fig. \ref{kf}c) with increasing doping can be attributed to the enhanced charge distribution imbalance between the OP and IP CuO$_2$ planes in Bi2223.

\begin{figure*}[tbp]
\begin{center}
\includegraphics[width=1\textwidth,angle=0]{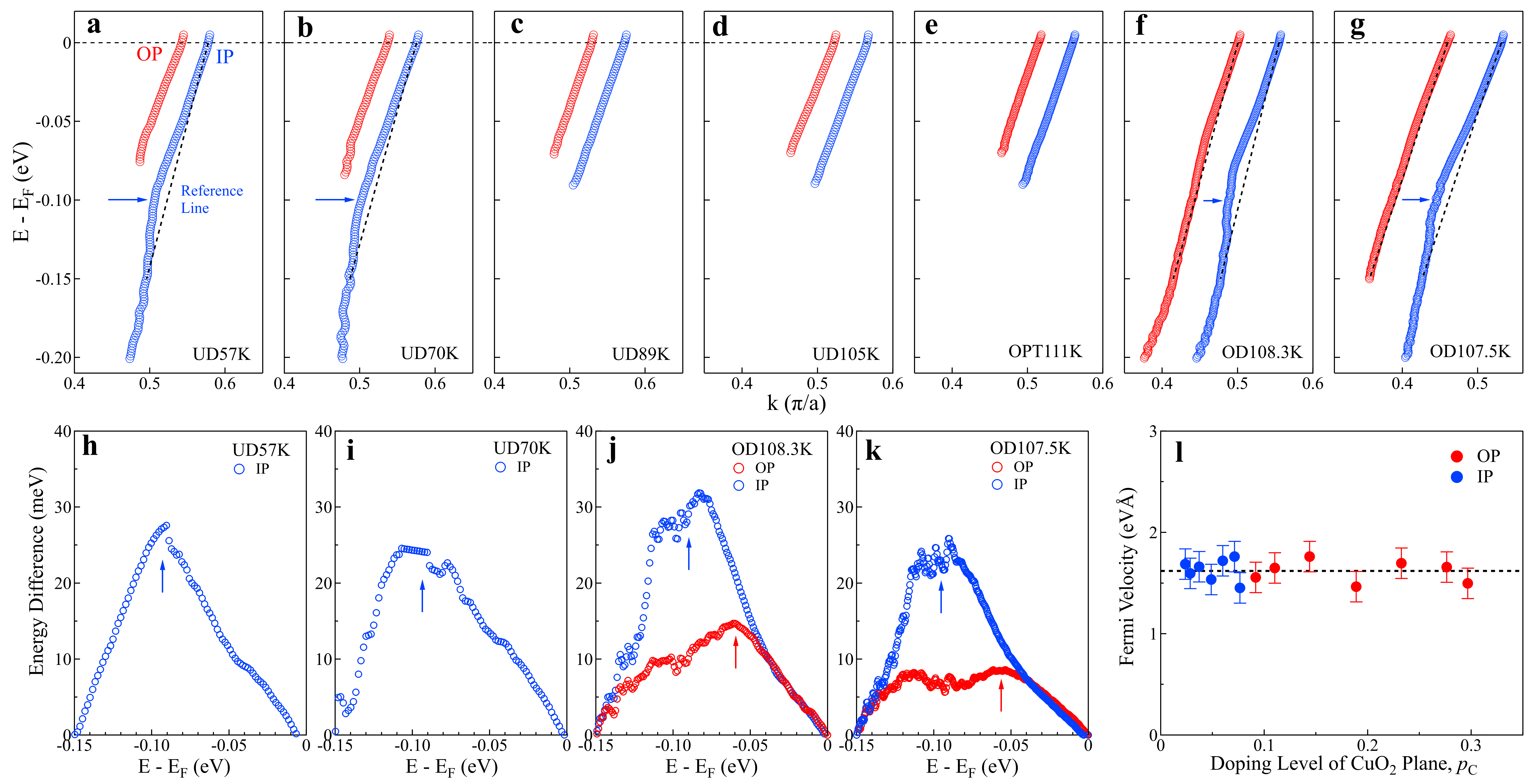}
\end{center}
\caption{{\bf Nodal electron dynamics in Bi2223 samples with different doping levels.}
\textbf{a-g}, MDC-derived nodal dispersions for the OP (red cirlces) and IP (blue cirlces) bands in Bi2223 samples with different doping levels. In \textbf{a, b, f, g}, straight lines are drawn by connecting two points in the related dispersions at the Fermi level and −0.15\,eV.
\textbf{h,i}, The energy difference for the IP bands in UD57\,K (\textbf{h}) and UD70\,K (\textbf{i}) Bi2223 samples, obtained by subtracting straight lines from the measured IP band dispersions in \textbf{a, b}.
\textbf{j,k}, The energy difference for the OP (red cirlces) and IP (blue cirlces) bands in OD108.3\,K and OD107.5\,K Bi2223 samples, obtained by subtracting straight lines from the measured OP and IP band dispersions in \textbf{f, g}.
\textbf{l}, Doping dependence of the Fermi velocity for the OP and IP bands in Bi2223. The Fermi velocity is obtained by fitting the measured OP and IP band dispersions in \textbf{a-g} within the energy range of [-40, 0]meV with a linear line. The dashed line shows a constant at 1.62\,eV\AA.
}
\label{Disp}
\end{figure*}

Now we come to analyze quantitatively the nodal band dispersions in Bi2223. Fig. \ref{Disp}a-g show dispersions of the OP and IP bands for Bi2223 with different doping levels, obtained by MDC fitting the band structures in Fig. \ref{Band}b. Considering the band intensity and the band overlapping, we only present the data that can be reliably extracted from the MDC fitting. Since the IP band in UD57K and UD70K samples is dominant (Fig. \ref{Band}b), its dispersion can be obtained up to a high binding energy as shown in Fig. \ref{Disp}a,b. For the heavily overdoped OD108.3K and OD107.5K samples, the OP and IP bands are well separated and clearly observed (Fig. \ref{Band}b) and their dispersions can also be extracted up to high binding energy, as shown in Fig. \ref{Disp}f,g. The IP band dispersions exhibit a clear kink at $\sim$100\,meV in both underdoped (Fig. \ref{Disp}a,b) and heavily overdoped (Fig. \ref{Disp}f,g) Bi2223 samples. The kink effect on the OP band dispersions in the heavily overdoped (Fig. \ref{Disp}f,g) samples is much weaker than that observed in IP band dispersions.

To identify the kink effect and determine the kink position, Fig. \ref{Disp}h,i show the energy difference between the measured IP band dispersion (blue circles in Fig. \ref{Disp}a,b) and the reference line (black dashed lines in Fig. \ref{Disp}a,b) in UD57K and UD70K samples. Fig. \ref{Disp}j,k show the energy difference between the measured OP (red circles in Fig. \ref{Disp}f,g) and IP (blue circles in Fig. \ref{Disp}f,g) band dispersions and their corresponding reference lines (black dashed lines in Fig. \ref{Disp}f,g) in OD108.3K and OD107.5K samples. The energy difference for the IP dispersions (blue circles in Fig. \ref{Disp}h-k) shows a peak or hump centered around 94\,meV (blue arrows in Fig. \ref{Disp}h-k) which is broad in overdoped samples. On the other hand, the energy difference for the OP dispersions (red circles in Fig. \ref{Disp}j,k) is much weaker than that for IP dispersions (blue circles in Fig. \ref{Disp}j,k). They show a weak peak at around 60\,meV (red arrows in Fig. \ref{Disp}j,k). The nodal dispersion kink has been commonly observed in single-layer and bilayer cuprates at $\sim$70\,meV\cite{bogdanov2000evidence, johnson2001doping, kaminski2001renormalization, lanzara2001evidence, zhou_universal_2003}. Further studies show that the 70\,meV kink coexists with another 40\,meV kink in the nodal dispersions\cite{yan_ubiquitous_2023}. The kink energy at $\sim$60\,meV in the OP nodal band dispersion is comparable to the $\sim$70\,meV observed before. However, the kink energy at $\sim$94\,meV we observed in the IP dispersion is significantly higher than those ($\sim$70\,meV and $\sim$40\,meV) observed before. The nodal dispersion kinks are generally attributed to electron coupling with bosons and specifically phonons\cite{bogdanov2000evidence, johnson2001doping, kaminski2001renormalization, lanzara2001evidence, zhou_universal_2003,yan_ubiquitous_2023}. It remains unclear whether the $\sim$70\,meV mode corresponds to vibrations of the apical oxygens or in-plane oxygens because their energy scales are similar. Our present study of the dispersion kink in the nodal IP band in Bi2223 may provide new clues to understand the origin of the $\sim$70\,meV kink. In single-layer and bilayer cuprates, the Cu in the CuO$_2$ plane is coordinated with the apical oxygens. But in Bi2223, the Cu in the inner CuO$_2$ plane is coordinated only with the four in-plane oxygens and no longer coordinated with the apical oxygens (Fig. \ref{Tc}a). The unusual kink effect we observed in the IP nodal band dispersion may be related to such a distinct Cu coordination configuration in Bi2223.

Figure \ref{Disp}l shows the Fermi velocity of the OP and IP bands in Bi2223 as a function of the doping level of the CuO$_2$ planes. The Fermi velocity is obtained by fitting the nodal OP and IP band dispersions in Fig. \ref{Disp}a-g within the energy range of [-40, 0]meV. The Fermi velocity of both the OP and IP bands show little variation with the doping level; it is nearly a constant at 1.62\,eV\AA. This observation of universal Fermi velocity is consistent with the previous results in single-layer and bilayer cuprates\cite{zhou_universal_2003}.

In summary, we investigated the doping-dependent nodal electron dynamics in the trilayer cuprate superconductor Bi2223 using high-resolution laser-based ARPES. By preparing samples spanning underdoped to overdoped regions, the electronic phase diagram is established. The nodal Fermi momentum for both outer (OP) and inner (IP) CuO$_2$ planes decreases monotonically with doping, while their momentum separation increases due to enhanced charge imbalance between the OP and IP planes. The IP band exhibits a prominent $\sim$94\,meV kink and a weak $\sim$60\,meV kink is observed in the OP band. The nodal Fermi velocity is nearly constant at  $\sim$1.62\,eV\AA\;across all doping levels. These results provide important information to understand the origin of high T$_c$ and superconductivity mechanism in high temperature cuprate superconductors.

\newpage


\vspace{3mm}


\vspace{3mm}

\noindent {\bf Acknowledgement}\\
 This work is supported by the National Natural Science Foundation of China (Grant Nos. 12488201 by X.J.Z., 12374066 by L.Z. and 12374154 by X.T.L.), the National Key Research and Development Program of China (Grant Nos. 2021YFA1401800 by X.J.Z., 2022YFA1604200 by L.Z., 2022YFA1403900 by G.D.L. and 2023YFA1406000 by X.T.L.), the Strategic Priority Research Program (B) of the Chinese Academy of Sciences (Grant No. XDB25000000 by X.J.Z.), Innovation Program for Quantum Science and Technology (Grant No. 2021ZD0301800 by X.J.Z.), the Youth Innovation Promotion Association of CAS (Grant No. Y2021006 by L.Z.) and the Synergetic Extreme Condition User Facility (SECUF).

\vspace{3mm}
\noindent {\bf Author Contributions}\\
 X.J.Z., L.Z., H.C. and J.M.S. proposed and designed the research. H.C. and J.M.S. carried out the ARPES experiments. C.T.L grew the Bi2223 single crystals. H.C. and J.M.S. contributed in sample preparation. X.Y.L., Y.W.C., C.H.Y, Y.J.S., J.X.Z., T.M.M, B.L., W.P.Z, N.C., X.L.R., S.J.Z, Z.M.W, F.F.Z., F.Y., Q.J.P., Z.Y.X., G.D.L., X.T.L., H.Q.M., L.Z. and X.J.Z. contributed to the development and maintenance of the ARPES systems and related software development. H.C., J.M.S., L.Z. and X.J.Z. analyzed the data. X.J.Z., L.Z., H.C. and J.M.S. wrote the paper. All authors participated in discussions and comments on the paper.



\clearpage
\begin{table}[!htbp]
  \caption{\enspace Preparation of underdoped Bi2223 samples with different T$_c$s by annealing the optimally-doped sample in vaccum at different temperatures for different durations.}
  \label{2223}
  \centering
  \footnotesize
  \setlength{\tabcolsep}{12pt}
  \renewcommand{\arraystretch}{1.5}
  \begin{tabular}{ccccccccc}
      \hline
      Annealing Temperature & \multicolumn{7}{c}{Annealing Duration, $\mathrm{T_c}$}\\
      \hline
      \multirow{2}{*}{500°C} & 1h & 4h & 4.5h & 5.5h & 8h & 10h & 33h\\
                            \cline{2-8}
                             & 108K & 88K & 88K & 69K & 60K & 30K & NSC\\
      \hline
      \multirow{2}{*}{450°C} & 15h & 24h & 48h & 41h & 72h &  & \\
                            \cline{2-8}
                             & 42K & 32K & 19K & NSC & NSC &  & \\
      \hline
      \multirow{2}{*}{420°C} & 15h & 18.5h & 72h & 72h & 96h &  & \\
                            \cline{2-8}
                             & 47K & 22K & 33K & 35K & 34K &  & \\
      \hline
      \multirow{2}{*}{400°C} & 15h & 24h & 81h &  &  &  & \\
                            \cline{2-8}
                             & 80K & 59K & 40K &  &  &  & \\
      \hline
  \end{tabular}
\end{table}

\newpage

\clearpage

\clearpage

\clearpage

\end{document}